\documentclass[12pt]{article}
\usepackage{caption}
\title{\vspace*{-48pt} A Geometric Framework for Odor Representation}
\author{Jack A. Cook and Thomas A. Cleland}
\date{Dept. Psychology, Cornell University, Ithaca, NY 14853, USA}
\usepackage{amssymb}
\usepackage{amsmath}
\usepackage{mathrsfs}
\usepackage{bbm}
\usepackage{graphicx}
\usepackage[margin=1 in]{geometry}
\usepackage{ragged2e}
\usepackage{palatino}
\usepackage{mathpazo}
\usepackage[utf8]{inputenc}
\usepackage[english]{babel}
\usepackage{amsthm}
\usepackage[square,sort&compress,comma,numbers]{natbib}
\usepackage{textcomp}
\def\style{naturemag}

\usepackage{changepage}

\usepackage{placeins}

\usepackage{pgffor}
\usepackage{asymptote}

\usepackage{tikz}
\usepackage{pgfplots}
\usepackage{verbatim}
\usetikzlibrary{arrows}
\usetikzlibrary{cd}
\usetikzlibrary{shapes,calc,positioning}
\usetikzlibrary{calc,intersections}
\usetikzlibrary{bending}
\usetikzlibrary{decorations.pathreplacing}

\newtheorem{theorem}{Theorem}[section]

\newtheorem{lemma}[theorem]{Lemma}
\theoremstyle{definition}
\newtheorem{definition}[theorem]{Definition}

\newtheorem{corollary}[theorem]{Corollary}

\newtheorem{remark}[theorem]{Remark}

\newcommand{\comp}{\circ}

\newcommand{\Res}{\text{Res}}

\newcommand{\R}{\mathbb{R}}

\newcommand{\Z}{\mathbb{Z}}

\pgfplotsset{compat=1.16}

\begin{document}
\maketitle 
\setcounter{section}{1}

\subsection*{Abstract}
We present a generalized theoretical framework for olfactory representation and plasticity, using the theory of smooth manifolds and sheaves to depict categorical odor learning via distributed neural computation. Beginning with the space of all possible inputs to the olfactory system, we develop a dynamic model for odor learning that culminates in a perceptual space in which categorical odor representations are hierarchically constructed through experience, exhibiting statistically appropriate consequential regions and clear relationships between the broader and narrower identities to which a given odor might be assigned.  The model reflects both the sampling-based physical similarity relationships among odorants, as observed in physiological receptor response profiles, and the acquired, learning-dependent perceptual similarity relationships among odors that can be measured behaviorally, and defines the relationship between them. Individual training and experience generates correspondingly more sophisticated odor identification capabilities. Because these odor representations are constructed from experience and depend on local, distributed plasticity mechanisms, geometries that fix curvature are insufficient to describe the capabilities of the system.  This generative framework also encompasses hypotheses explaining representational drift in postbulbar circuits and the context-dependent remapping of perceptual similarity relationships.

\vspace{0.5cm}
\textbf{Keywords:}  set theory, contrast models, representational drift, olfactory bulb, category learning, topological spaces, geometry

\subsection*{Introduction}

The task of sensory systems is to provide organisms with reliable, actionable information about their environments.  However, useful information is not readily available; the environmental features that are ecologically relevant to an organism are rarely directly evident in primary receptor activation patterns.  Rather, these representations of interest must be \textit{constructed} from the combined signals of populations of sensory receptors.  This construction process is mediated by sophisticated networks of neural circuitry that draw out different aspects of potentially important information from raw sensory input patterns.  (We previously have proposed that these interactions and transformations are most effectively modeled as a \textit{cascade of successive representations} \cite{Cleland2014,Cleland2020Engin}, in which each neuronal ensemble constructs its representation by selectively sampling the activity of its antecedents).  With sufficient understanding of a given sensory modality, and of the neuronal architecture of the corresponding sensory system, these representations and their transformations can be geometrically modeled. 

Importantly, such vetted geometries sharply constrain how sensory representations can be physically encoded and transformed by neuronal circuitry.  For example, the high dimensionality of odorant similarity space during sampling establishes that contrast enhancement transformations operating in this similarity space \cite{Yokoi1995} cannot be mediated by nearest-neighbor lateral inhibition, as once was believed, but instead require novel circuit mechanisms within the olfactory bulb that are able to embed and transform high-dimensional features \cite{Cleland2006, Fukunaga2014}. That is, these fundamentally mathematical constraints concretely govern the architecture and function of neural circuits in the brain.  Establishing the geometries of representational cascades is foundational to understanding how the underlying neural systems compute.  

The representational cascade that underlies odor recognition and identification is impressively powerful and compact.  Olfactory bulb circuits (Fig. 1) impose an internally generated temporal structure on afferent inputs \cite{Werth2022,Li_Cleland2017,Li_Cleland2013,Kashiwadani1999,Bathellier2006,Eeckman1990} while also regulating contrast \cite{Cleland2006,Fukunaga2014}, normalizing neuronal activity levels \cite{Cleland2007,Cleland2011,Banerjee2015,Cleland2020Engin,Storace2017,Storace2019}, governing interareal communication \cite{Frederick2016,Kay2014}, and managing patterns of synaptic and structural plasticity \cite{Chatterjee2016,Strowbridge2009,Gao2009,Sailor2016,Lepousez2014,Gheusi2014,Magavi2005}.  The resulting perceptual system learns rapidly and is conspicuously resistant to retroactive and compound interference \cite{Herz1996, Stevenson2007}, as well as to interference from simultaneously encountered competing odorants (modeled in \cite{Imam2020}), which can profoundly degrade the odorant-specific receptor activity profiles upon which odor recognition ostensibly depends \cite{Gronowitz2021,Xu2020,Zak2020,Pfister2020,Inagaki2020}. These capacities accentuate the implications of the profound plasticity of the early olfactory system:  odor representations, and the basic process of olfactory perception itself, are fundamentally and critically dependent on learning \cite{Wilson2003,WilsonBook2006,Royet2013}.  Discrete, meaningful odors and their implications -- excepting a few species-specific innately recognizable odors -- must be categorically learned through individual experience.  Indeed, there is abundant evidence for the perceptual learning of meaningful odor representations, from their generalization properties \cite{Cleland2009,Cleland2011}, to the mechanisms of odor learning and memory \cite{Tong2014,Vinera2015,Wilson2003,Mandairon2011,Kermen2010,Grelat2018,Gheusi2014}, to the association of odor representations with meaning and context, even in peripheral networks such as the olfactory bulb \cite{Doucette2008,Nunez2014,Ramirez2018,WilsonBook2006,Mandairon2014,Herz2005,Aqrabawi2018a,Aqrabawi2020,Levinson2020,Li2020}.  What we lack is a common theoretical framework in which these diverse phenomena can be usefully embedded.

\begin{figure}
\vspace{0.5cm}
\centering  
  \includegraphics[width=6in]{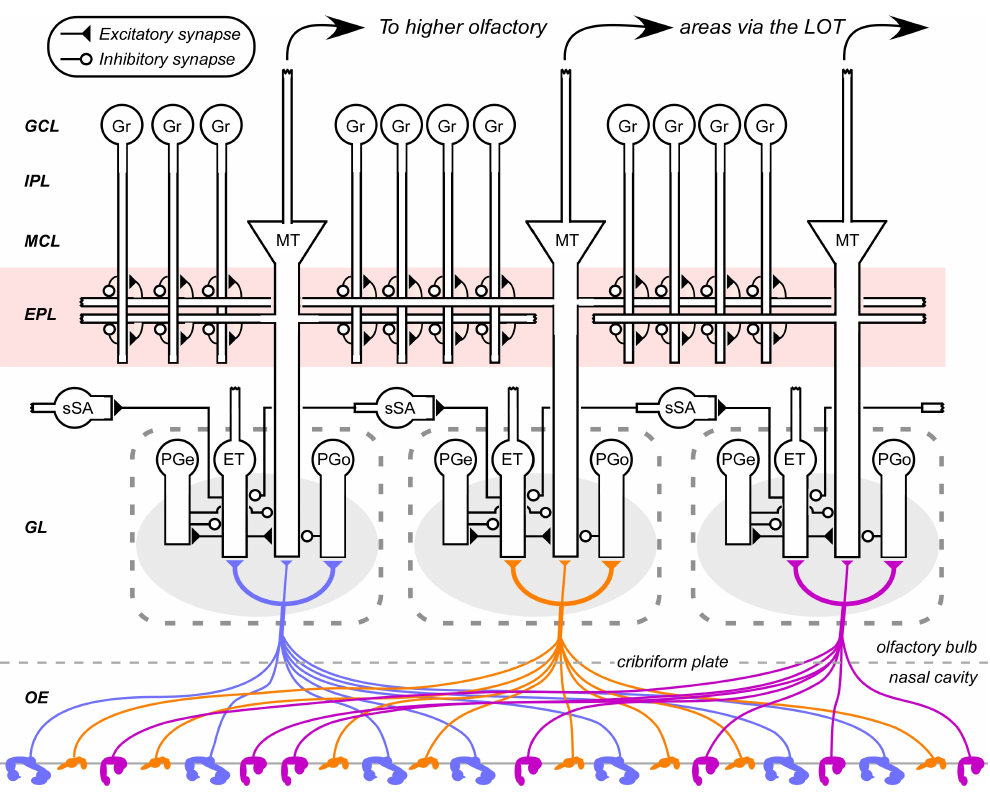}
  \vspace{0.25cm}
  \caption{{\bf Circuit diagram of the mammalian olfactory bulb.} The axons of primary olfactory sensory neurons expressing the same odorant receptor type converge together as they cross into the brain and arborize together to form \textit{glomeruli} (shaded ovals) across the surface of the olfactory bulb. Several classes of olfactory bulb neuron innervate each glomerulus, including external tufted cells (ET), olfactory nerve-driven periglomerular cells (PGo), and external tufted cell-driven periglomerular cells (PGe). Superficial short-axon cells (sSA) project broadly and laterally within the deep glomerular layer, interacting with glomerular interneurons and mediating global feedback inhibition \cite{Banerjee2015}. Principal neurons (MT) include mitral cells and projecting tufted cells, which interact with the dendrites of inhibitory granule cells (Gr) via reciprocal connections in the external plexiform layer (EPL), and project to several regions of the brain.  (Specifically, projecting tufted cells receive similar afferent input and interact with granule cells as do mitral cells, though their physiological responses and extrabulbar projection patterns differ \cite{Igarashi2012}. Whereas this geometric framework may apply comparably to both cell types, we refer herein to mitral cells for simplicity).  EPL interneurons and the multiple classes of deep short-axon cell are not depicted.  OE, olfactory epithelium (in the nasal cavity); GL, glomerular layer; MCL, mitral cell layer; IPL, internal plexiform layer; GCL, granule cell layer. Filled triangles denote excitatory synapses; open circles denote inhibitory synapses.  Adapted from \cite{Cleland2020Engin}.}  
 \vspace{0.5cm}
  \label{OBnet}
\end{figure}

\subsection*{Perceptual frameworks}
Theoretical frameworks for understanding sensory systems include \textit{perceptual spaces} and \textit{hierarchical structures}\footnote{Also see \textit{Semantic similarity and contrast models} section, below}.  Both are founded on metrics of similarity \cite{Zaidi2013,Edelman2012,Shepard1987,Clapper2019,Tversky1977}, though the former presumes an essentially continuous theoretical space of some dimensionality into which individual stimulus representations are deployed, whereas the latter presumes some degree of qualitative category membership for each such representation, with intercategory similarities mapping to the hierarchical proximities among categories.  Perceptual spaces can be defined  using a variety of metrics, including both \textit{physical metrics} such as wavelength (color) or frequency (pitch) and \textit{perceptual metrics} such as those revealed by generalization gradients \cite{Shepard1987,Cleland2009,Cleland2002} or by ratings on continuous scales by test subjects \cite{Khan2007}.  Indeed, study of the transformations between physical and perceptual metric spaces is foundational to understanding sensory systems from this perspective \cite{Zaidi2013,Meister2015,Victor2017}.  In contrast, hierarchical structures arise from perceptual categorization processes, though relationships among the resulting categories still may correlate with underlying similarities in the physical properties of stimuli. Critically, it is categories that are generally considered to be embedded with associative meaning (\textit{categorical perception}) \cite{Harnad1987,Goldstone2010,Aschauer2018,Locatelli2016}.  Consequently, a useful theoretical framework for olfactory perception must consider the construction of these categorical representations with respect to the physical similarity spaces that are sampled during sensory activity.  That is, along their representational cascades, olfactory sensory systems can be effectively considered to transition from a physical similarity space metric to a modified perceptual space, within which hierarchical category representations, generally corresponding to ecologically relevant odor sources, can be constructed. 

Interestingly, the olfactory modality lacks a clear, organism-independent physical metric such as wavelength or frequency along which the receptive fields of different sensory neuron populations can be deployed (and against which the nonuniform sampling properties of the sensory system can be measured) \cite{Cleland2014,Cleland2020Engin}.  However, olfaction does provide an objective basis for an \textit{organism-dependent} physical similarity space.  In this framework, the activity of each odorant receptor (OR) type -- e.g., each of the $\sim$400 different ORs of the human nose or the \textgreater1100 different ORs of the rodent nose -- comprises a distinct unit dimension.  Specifically, the instantaneous activation level of the convergent population of primary sensory neurons expressing each OR type \cite{Mombaerts1996,Cleland2020Engin} provides a value from zero to one (maximum activation), such that any possible olfactory stimulus can be mapped to a vector embedded in a physical metric space with dimensionality equal to the number of OR types. We refer to this receptor activation-based metric space as $R$-space \cite{Gronowitz2021} (Fig. 2A), within which individual vectors directly correspond to physiological measurements of olfactory sensory neuron activity (e.g., optical recordings from sensory neuron axonal arbors in the olfactory bulb glomerular layer \cite{Wachowiak2001,Storace2017,Storace2019}) and within-category variance is incorporated into the definition of a learned odor.  Critically, in this framework, (1) the dimensions of $R$-space are linearly independent of one another, and (2) every possible instantaneous profile of OR activation, including any occluding effects of multiple agonists and antagonists competing for common receptors, is interpretable and maps to a vector in $R$. \cite{Gronowitz2021}.  

\begin{figure}[!ht]
  \vspace{0.5cm}
  \centering  
  \includegraphics[width=6in]{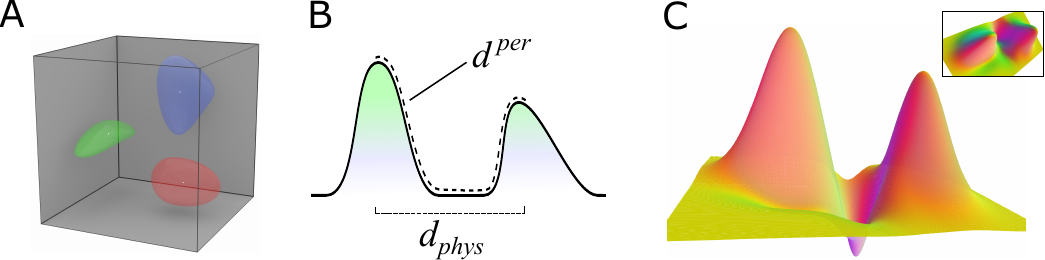}
  \vspace{0.5cm}
  \caption{{\bf Depictions of olfactory spaces.  [A]} Three-dimensional $R$-space containing three odor source volumes. Each of the odors activates all of the three receptors (axes) depicted, but with different ratios of activation.  \textbf{[B]} One-dimensional depiction of $S$-space containing two learned odors, illustrating the relationship between the physical (\textit{d\textsubscript{phys}}; distance along abscissa) and perceptual (\textit{d\textsuperscript{per}}; arc length along the surface) measures of similarity between these separately learned odor representations.  Between-category separation renders $\textit{d\textsuperscript{per}} > \textit{d\textsubscript{phys}}$. The odor representations need not be symmetrical, similar in shape to one another, or equivalent in peak height or area under the distension.  Their shapes will be directly reflected in perceptual generalization gradients \cite{Cleland2009, Cleland2011}. \textbf{[C]} Two-dimensional depiction of $S$-space with learning-based distensions into the third dimension. The two odor representations depicted are based on two different ratios of activation of the two receptors (axes) depicted.  The odors are physically similar (proximal), but have been differentiated both by their statistical learning as separate odors (as in panel B) and additionally by specific interrepresentational discrimination learning (negative distension located specifically between the two odor representations, further increasing the perceptual difference \textit{d\textsuperscript{per}} between these odors). In higher-dimensional spaces, two such odor representations can be separated to a nearly arbitrary degree without affecting similarity relationships among other nearby odor representations. This can lead to violations of the triangle inequality in \textit{d\textsuperscript{per}} (i.e., $\overline{ACB} < \overline{AB}$).   \textit{Inset.} Top view of the same two odor representations as in panel C. Note that the variances of the two receptor dimensions of each representation are independent, and not necessarily symmetrical.}
  \vspace{0.5cm}
  \hrule
  \label{Landscapes}
\end{figure}

Linear independence among the dimensions of $R$-space is important for analytical purposes, but their orthogonality is irrelevant \cite{Cooperstein2015}\footnote{Specifically, by the Gram-Schmidt process, the collection of linearly independent sets of maximal size is in correspondence with the collection of orthonormal bases.}.  This is a vital distinction, not least because the orthogonality of OR receptive fields depends on the properties of the chemosensory input space -- that is, the chemical environment -- and hence cannot be uniquely defined as a property of the olfactory system \textit{per se}.  In principle, each OR type should have regions of its receptive field that distinguish it from any other single OR type, such that activation of a given OR need not always imply activation of a particular different OR (that is, no two dimensions will be identical)\footnote{If we think of receptors as functions, then this is saying that the collection of receptors separates points.}.  However, within any given sensory world, as defined by a finite set of odorant stimuli with established properties and probabilities of encounter, there will be reliable activity correlations among many pairs of receptor types that can support substantial dimensionality reduction \cite{Haddad2010} (see \textit{Efficient coding} section).  Critically, however, these reduced dimensionalities are not characteristic of the olfactory system \textit{per se} at this level, as they are strongly reflective of the statistics of the stimulus set used and its particular interactions with the deployed complement of receptors. 

Similar dimensionality-reduction efforts also have been applied to olfactory perceptual data \cite{Koulakov2011,Castro2013,Zarzo2006}.  These results also are substantially determined by the particular sets of stimuli employed, but additionally engage the problem of just what the olfactory system constructs from the space of its underlying physical inputs.  It is reasonably clear (even axiomatic) that the  sampling of physical odorant spaces is highly nonuniform \cite{Castro2013,Koulakov2011} -- that is, odor samples are \textit{signal sparse} \cite{Berke2017} -- but, perhaps more importantly, the process of perceptual learning itself directly affects perceived odor similarity relationships, as can be measured with generalization gradients \cite{Cleland2009,Cleland2011,Perez2016,Wright2008,Daly2001}.  A general framework for olfactory perception must reflect all of these phenomena, embedding physical and perceptual similarity spaces into a common geometric framework that admits the construction of experience-dependent perceptual categories.  

\subsection*{The geometries of olfaction}
In addition to dimensionality, the second fundamental property of a sensory space is its intrinsic geometry \cite{Zaidi2013}.  Establishing a geometry provides access to theorems by which representational structures can be formally defined and manipulated. However, it is not necessary to restrict the topology of a sensory space to a single geometry with fixed curvature -- in fact, as indicated below, this is neither advisable nor ultimately possible for the olfactory modality.  Specifically, we show that a mature olfactory perceptual space cannot be simply characterized as "Euclidean", "spherical", "hyperbolic", or otherwise, as its dependence on localized, experience-dependent plasticity instead produces a space comprising locally modifiable regions that can exhibit the properties of different geometries.  These can be glued together formally via the theory of sheaves.  

We present a generalized geometric framework for the construction of odor representations. The framework is based on the molecular/physiological encoding capacities of the olfactory bulb input layer \cite{Gronowitz2021} and the transformation of these physiological \textit{odorant} representations by perceptual learning into cognitive, categorical \textit{odor} representations to which meaning can be ascribed.  Key features of this framework include the simultaneous depiction of sampling-based physical similarity and learning-dependent perceptual similarity within the perceptual space, and a categorization process that culminates in a perceptual space within which qualitatively discrete \textit{odor} representations are hierarchically constructed through experience, exhibiting statistically appropriate consequential regions with probabilistic boundaries that reflect learned generalization gradients \cite{Cleland2009,Cleland2011,Shepard1987,Locatelli2016}.  Critically, individual training and experience generates progressively more sophisticated hierarchies and concomitantly superior odor identification capabilities \cite{Royet2013,Rabin1988}.  

A simplified illustration of the analytical framework is depicted in Diagram 1. Briefly, the space of instantaneous physical inputs to an olfactory receptor activation space ($R$-space) comprising $N$ receptor types can be depicted as an $N$-dimensional unit cube (Fig. 2A).  Transformations arising  from initial glomerular-layer computations \cite{Cleland2020Engin} generate a modified receptor space termed $R'$; this space inherits the dimensionality of $R$-space but respects the nonuniform likelihoods of different state points within that space.  The subsequent transformation from $R'$ to $S$-space ("scent space") reflects the perceptual learning processes that construct categorical representations of meaningful \textit{odors}.  

This theoretical model does not depend on precisely where in the olfactory representational cascade these transformations occur.  However, we consider that the map $B$ from $R$-space to $R'$-space reflects signal conditioning computations performed within the glomerular layer of the olfactory bulb \cite{Cleland2014,Cleland2020Engin}, whereas the subsequent transformation into $S$-space is mediated by computations within the olfactory bulb external plexiform layer network, inclusive of its reciprocal interactions with deeper olfactory structures (Fig. 1).  Briefly, we propose that the construction of categorical odor representations through statistical experience arises from learning-dependent weight changes between mitral cell principal neurons and granule cell interneurons in the external plexiform layer of the olfactory bulb \cite{Imam2020}, inclusive of the process of adult neurogenesis, which integrates new interneurons into the OB circuit via an experience-dependent mechanism and is necessary for odor learning \cite{Moreno2009,Sailor2016,Lepousez2014,Gheusi2014,Grelat2018,Forest2019,Kermen2010,Sultan2010}. The configural receptive fields of granule cells (as proposed in \cite{Imam2020,Cleland2020Engin}) provide a high-dimensional state space that enables the representation and storage of statistical priors \cite{Singer2016}. To construct this theoretical $S$-space, and attribute to it the capacities of generalization and experience-dependent hierarchical categorization, we first build a transitional space $M$ based on mitral cell activity representations, inclusive of the actions performed on these representations via their interactions with granule cell interneurons:  

\vspace{0.5cm}
\begin{equation}\label{Diagram1} 
\begin{tikzcd}
R\arrow[r,"B"] &R' \arrow[dr,"\xi"] \arrow[d,swap,"\Delta"] \\
&S& M \arrow[l,"h"] 
\end{tikzcd} 
\end{equation}
\text{}
with $h\in C^\infty(\R^m).$ 

Formally, $R$ is a unit parallelepiped defined by primary olfactory receptor activation levels. $R'$ denotes a subspace of normalized points, following the transformation of sensory input by glomerular-layer circuitry, and is the image $B(R).$  $M$ is a vector bundle over $R'$ of rank equal to the number of mitral cells (Fig. 1) and is generated by mitral cell output. $\xi$ denotes the input presented to mitral cells following glomerular processing, comprising a sparsened, statistically conservative manifold \cite{Borthakur2019}; it is a section of the vector bundle $M$. $S$ denotes the perceptual space, and is realized as a transformation $\Delta$ of $R'$-space that embeds odor learning via smooth functions.   This resulting $S$-space does not, indeed cannot, admit a single geometry, because of the essential requirement for locally adaptable curvature.  We describe this generative process in detail below. 

\setcounter{theorem}{0}
\subsection*{$R$-Space}
The first representational space in olfaction is directly derived from the ligands of the physical odorant stimulus interacting with the set of chemoreceptive fields presented by the animal's primary odorant receptor complement \cite{Gronowitz2021}.  Both vertebrate and arthropod olfactory systems are based on large numbers of receptor neurons, each of which expresses one primary odorant receptor out of a family of tens (in \textit{Drosophila}) to over 1000 (in mice, rats, and dogs).  The axons of primary sensory neurons expressing the same receptor converge together to form discrete \textit{glomeruli} across the surface of the olfactory bulb (in vertebrates; the arthropod analogue is the antennal lobe), enabling second-order principal neurons (e.g., mitral cells) to sample selectively from one or a few receptor types (depending on species; see \textit{$M$-Space} section).  The response of each receptor type to an odor stimulus constitutes a unit vector that can range in magnitude from nonresponsive (0) to maximally activated (i.e., receptor saturated with agonist; 1).  A complete representational space for instantaneous samples of this input stream consequently has a dimensionality equal to the number of odorant receptor types $N$.  That is, in a species with three odorant receptors, the space containing all possible instantaneous input signals would be a three-dimensional unit parallelepiped (Fig. 2A), whereas the $R$-space of a mouse expressing 1100 receptor types would comprise a 1100-dimensional unit parallelepiped.  Notably, odorant representations in $R$-space can be directly measured via optical recordings of activity in the receptor neuron axonal arbors of olfactory bulb glomeruli \cite{Wachowiak2001,Storace2017,Storace2019}, though to specify an odorant vector completely would require activity measurements from every glomerulus. Because the vector coordinate in each dimension depends on the activation of a qualitatively distinct odorant receptor type, the dimensions can be considered independent of one another.  For this, as noted above, it is not necessary that the receptive fields of the different ORs be  orthogonal under all conditions, only that they be linearly independent \cite{Cooperstein2015}; indeed, the orthogonality of their response vectors cannot even be defined without reference to the statistics of the particular physical environment in which they are deployed.  

Formally, $R$-space is defined as the space of linear combinations of these vectors with coefficients in $(0,1).$ Consider the space of all possible odorant stimuli in a species expressing $N$ odorant receptor classes. Each odorant stimulus $s^*$ corresponds to a unique instantaneous glomerular response profile that can be represented as a vector $s^*\in \R^N$.  Normalizing the activation in each glomerulus enables us to consider $s^*\in \prod^n(0,1)$, the unit cube in $N$ dimensions. Denote this receptor activation-based representational space $R$. As $R$ is open in the ambient space $\R^N$, $R$ also has dimension $N$ as a manifold.  

By considering a product of spaces, we are assuming that the responses of different glomeruli are orthogonal. In the greatest generality, we would need to consider points on a unit parallelepiped generated by the primary receptors. That said, we can apply an invertible linear transformation (specifically, the matrix generated by the Gram-Schmidt process \cite{Cooperstein2015}) to this parallelepiped to generate a cube -- a mathematical formalism that does not affect the particulars of this situation. Consequently, for  the  remaining sections, we can assume without a loss  of  generality that $R=\prod^n(0,1).$  

\subsection*{Glomerular-layer computations, $R'$}
The first computational layer of the olfactory bulb -- the glomerular layer -- computes a number of transformations important for the integrity and utility of odor representations, including contrast enhancement \cite{Cleland2006}, global normalization underlying concentration tolerance \cite{Cleland2011,Banerjee2015}, and potentially other effects \cite{Cleland2020Engin}.  These processes substantially alter the respective probabilities of the points in $R$-space.  For example, global feedback normalization in the deep glomerular layer \cite{Banerjee2015} ensures that points at which most or all of the vectors have very high values will be improbable.  The outcome of this transformation is represented as $R'$, essentially a smooth manifold embedded in $R$-space. 

In addition to the systematically unlikely points in $R$ that are omitted from the manifold $R'$, it is also the case that, under natural circumstances, most of the possible sensory stimuli $s^*$ that could be encountered in $R'$ actually never will be encountered in an organism's lifetime.  That is, odor representations within $R$-space are \textit{signal sparse} \cite{Berke2017}. Moreover, we argue that \textit{odor sources} $s^*$ are discrete, but inclusive of characteristic variance in quality, and hence constitute \textit{volumes} (manifolds) within $R'$. To account for this, we denote this variance by $s^*=(x,U_x)$, where $x\in R'$ and $U_x$ denotes an $n$-tuple of variances (i.e., one variance for each dimension of freedom in $R'$). That is to say, \[U_x=(\sigma_1^2,...,\sigma_n^2)\]
From this we arrive at the following definition:

\begin{definition}
    A pair $(x,U_x)$ constitutes an \textit{odor source volume} in $R'$ if $U_x$ is a non-empty simply connected neighborhood of $x$ and $(x,U_x)=s^*$ for some odorant $s^*.$  
\end{definition}

\noindent
That is, an odor source volume corresponds to a manifold within $R'$ that comprises the population of odorant stimulus vectors arising from the range of variance in receptor activation patterns exhibited by a particular, potentially meaningful, odor source (Fig. 2A) -- here defining \textit{source} as the odor of some thing (i.e., the \textit{sign-vehicle} \cite{Bruni2008}), as opposed to the thing that smells (the \textit{sign-object}).  This includes variance arising from nonlinearities in concentration tolerance mechanisms that cannot be fully avoided \cite{Cleland2011} as well as genuine quality variance across different examples of a source.  (For example, the odors of \textit{oranges} vary across cultivars and degrees of ripeness; the odors of \textit{red wines} vary across grape cultivars, terroir, and production methods).  The source volume in $R'$ thereby corresponds to an odor source (e.g., orange, red wine), inclusive of its variance, and delineates the consequential region of the corresponding \textit{odor} category that will be developed via perceptual learning.  Critically, it is not important at this stage to specify multiple levels of organization within odor sources (e.g., red wine, resolved into Malbec, Cabernet, Montepulciano, etc., then resolved further by producer and season); it is the process of odor learning itself that will progressively construct this hierarchy of representations at a level of sophistication corresponding to individual training and experience. 

\subsection*{$M$-Space}
The transformation from $R'$ to $S$-space depicted in Diagram 1 is mediated by the interactions of mitral and granule cells.  In this framework, mitral cells directly inherit afferent glomerular activity from $R'$ (Diagram 1, $\xi$), but their activity also is modified substantially by patterns of granule cell inhibition that, via experience-dependent plasticity, effectively modify mitral cell receptive fields to also incorporate higher-order statistical dependencies sourced from the entire multiglomerular field.  (A simplified computational implementation of this constructive plasticity is presented in the learning rules of \cite{Imam2020}).  This is depicted in Diagram 1 as an effect $C^\infty(\R^m)$ of a mitral cell product space $M$ which contributes to the construction of $S$, in order to highlight the smooth deformations of $R'$ into $S$ via passage to $M.$    

These effects of mitral cell interactions, arising from experience, are modeled locally as a product space $M$ based on the principle that each glomerulus -- corresponding to a receptor type in $R'$ -- directly contributes to the activity of some number of distinct mitral cells.  In the mammalian architecture, mitral cells receive direct afferent input from only a single glomerulus, such that the afferent activation of each mitral cell (or group of sibling mitral cells) corresponds directly to a single receptor type.  In this special case, $M$-space is globally a product.  To formalize this, we label the glomeruli $g_1,...,g_q.$ To each glomerulus, we associate the number of mitral cells that sample from it; denoted $m_i\in \Z.$ Let $k=\sum^q m_i.$ Then, the naive space constructed from these data is \[R'\times \R^k=\{(r,v):r\in R',v\in \R^k\}     \]  
The interpretation of this space is as follows. To each point in $R'$, we can associate a vector that is an identifier for how subsequent mitral-granule cell interactions in the olfactory bulb will transform the input in service to identifying it as a known percept.  The manifolds associated with particular odor source volumes in $R'$ will, owing to experience-dependent plasticity, come to exhibit related vectors that, in concert, manifest source-associated consequential regions \cite{Shepard1987}.  These regions reflect categorical perceptual representations \cite{Locatelli2016} and are measurable as olfactory generalization gradients \cite{Cleland2009,Daly2001}. That is, the learning-dependent plasticity of synaptic interactions in the OB underlies the creation of categorical odor representations and manages between-category perceptual separation.  Simplified computational implementations have depicted these acquired representations as fixed-point attractors, tolerant of background interference and sampling error but lacking explicit consequential regions \cite{Imam2020}.

$M$ is always globally a product space (as $R'$ is contractible). For the mammalian architecture, the dimensionality $m$ of mitral cell output (grouping sibling mitral cells together) is identical to that of glomerular output $k$.  In nonmammalian tetrapods, in contrast, individual mitral cells may sample from more than one glomerulus \cite{Mori81a,Mori81b}.  This introduces a reduction (in general) of the product space such that $m$ now can be less than $k$.  In the general case where $m \leq k$, the mitral cell space becomes a real rank $m$ vector bundle \[\R^m\hookrightarrow M \overset{\pi}{\rightarrow} R'\] over $R'$. In the mammalian architecture, because $m=k$,  \[M=R'\times \R^m\] 
thus rendering $M$ a smooth manifold with the convenient property that to every input $x\in R'$ we associate a point $(x,v)$, where $v$ is a vector whose $i^{th}$ component is the value of the output of the $i^{th}$ mitral cell. 

Formally, $M$ is a (trivial) vector bundle over $R'$ with fibre $\R^m.$ Then, the smooth maps that send $x\mapsto (x,v)$, such that composition with projection onto $R'$ is the identity, are called global smooth sections of the bundle, and the set of these is denoted $\Gamma(R',M).$ To any smooth manifold $P$, we can associate a sheaf (of rings) of smooth functions \[C^\infty(P)=\{f:P\to \R:f \text{ is smooth}\}\] To any open subset, we have a restriction map $\Res^P_U:C^\infty(P)\to C^\infty(U).$ In general, if $U\subseteq P$ is open, then $\Gamma(U,E)$ is a $C^\infty(U)-$module for any bundle $\pi:E\to P.$  $C^\infty(-)$ makes $P$ into a locally ringed space and $\Gamma(-,E)$ is a sheaf of $C^\infty(-)$-modules. These two sheaves, $C^\infty(P)$ and $\Gamma(-,E)$, fully characterize all of the properties of the manifold and vector bundle, and, critically, \textit{enable the geometric representation of localized synaptic plasticity by establishing an algebraic interpretation of local information that can be combined formally into global information.} (The theory of sheaves is described in more detail below). Precisely, to any open cover $\{U_i\}$ of $R'$ and a collection of elements $s_i\in C^\infty(U_i)$ such that $s_i|_{U_i\cap U_j}=s_j|_{U_i\cap U_j}$ for all $i,j$ there exists a unique global lift $s\in C^\infty(R')$ such that $s|_{U_i}=s_i$. By this method, locally-defined information arising from synaptic plasticity can be glued together into a coherent whole.\footnote{
This presents the question: what mathematically distinguishes these mammalian and nonmammalian architectures if their geometry and topology are essentially equivalent? One answer is as follows. In the mammalian architecture, every odorant induces $m$ distinct functions $f_1,...,f_m$ that depend only on a single coordinate of the input odorant and yield the coordinates of the corresponding mitral cell activation level, essentially rendering them as maps $f_i:\R\to \R$. That is to say, tracing an odorant $x=(x_1,...,x_N)$ through the diagram says that we associate to it a pair $(x,v)$ where \[v=[f_1(x_1),...,f_{m_1}(x_1),f_{m_1+1}(x_2),...,f_{m_1+m_2}(x_2),...,f_{m}(x_N)]^T\] In contrast, in the non-mammalian case, the functions generated by sensory input depend on more than one input coordinate. By labeling the mitral cells using the integers $1,...,m$ and setting \[ e(i):=\{\#\; \text{of distinct glomerular inputs}\} \] these functions become maps $f_i:\R^{e(i)}\to \R.$  While the mappings are foundationally similar, this difference renders the latter maps more complicated to analyze.}  

\subsection*{$S$-Space}
$S$-space, or scent space, is a constructed perceptual space tasked with preserving physical similarity relationships among odorants while also embedding the transformations arising from perceptual learning, specifically including those forming incipient categorical \textit{odors}.  To do this, we embed $R'$ into a higher-dimensional space $S$ (with dimension $N+1$).  Under this embedding, we represent perceptual learning in $S$ by growing $U_x$ in the positive $N+1th$ direction around odor source volumes in $R'$. (Discrimination training also can grow $U_x$ in the negative $N+1th$ direction, as discussed below). This transformation does not affect distance relationships in ${\R^N}$, but systematically increases them in ${\R^{N+1}}$, reflecting the process of \textit{between-category separation} \cite{PGJ-Harnad2019,Goldstone2010}.  To quantify this transformation, we construct two distance metrics, $d_{phys}$ and $d^{per}$, on $S$ (Fig. 2B).

\begin{definition}
	Let $x,y\in S$ be two points.  We define the \textit{physical metric} between the two points as the Euclidean distance between them in $R.$ In notation, \[d_{phys}(x,y)=|\pi_{\R^N}(x)-\pi_{\R^N}(y)|  \] 
\end{definition}
\noindent This metric reflects the physical similarities between odorants in the receptor space, as defined by commonalities in receptor activation profiles, and which are not affected by perceptual learning.  
\begin{definition}
	Let $x,y\in S.$ Consider $x$ and $y$ as vectors in $\R^{N+1}$. Then, let $\gamma:[0,1]\to S$ be the curve defined by $\gamma(0)=x$, $\gamma(1)=y$ and $\pi_{\R^{N}}(\gamma'(t))=w\cdot\left[  \pi_{\R^N}(\gamma(1))-\pi_{\R^N}(\gamma(0))\right]$ with $w$ some real number dependent on $t$. The \textit{perceptual metric}, \[d^{per}(x,y)=\int_0^1||\gamma'(t)||dt  \] is the arc-length along the surface of $S$ between the points $x$ and $y$, and reflects the acquired semantic similarities and differences among odors \cite{Hahn2001} (Fig. 2B). Note that $\pi_{\R^N}(\gamma')$ is well defined as $S\hookrightarrow \R^{N+1}$ and thus the tangent space $T_{\gamma(t)}S\subseteq T_{\gamma(t)}\R^{N+1}=\R^{N+1}.$ 
\end{definition}  

The relationship between these two metrics tracks the changes in $S$ induced by the construction of odor representations; specifically, $d^{per}$ reflects experience-dependent changes in the perceptual distance between $x,y\in S$ that are excluded from the $d_{phys}$ metric (Fig. 2B). Learning about an odor source $(x,U_x)$ progressively distends the volume (in $\R^N$) in the $N+1th$ direction, such that the shape of this distension gradually will come to reflect the odor source volume in $R'$.  That is, over time, the breadths (in each of the $N$ dimensions) of the distension into the additional ($N+1th$) dimension will come to reflect the actual variances $U_x$ of the odor source $s^*=(x,U_x)$ as naturally encountered \cite{Cleland2011}.  The quasi-discrete distensions formed in the $N+1th$ dimension correspond to incipient categories -- i.e., categorically perceived \textit{odors} -- and their breadths and gradients can be measured behaviorally as generalization gradients \cite{Cleland2009,Cleland2011}. Importantly, the variance for each dimension of freedom of $U_x=(\sigma_1^2,...,\sigma_n^2)$ in $R'$ is independent; that is, different samples of a given natural odor source may vary substantially in some aspects of quality but not others, where an aspect of quality refers to the relative levels of activation of a given odorant receptor type (Fig. 2C). 

Formally, to construct the perceptual space $S$ such that there exists a perceptual metric $d^{per}$ that interacts with the natural physical metric $d_{phys}$ of $R'$, we consider the embedding $R'\hookrightarrow \R^{N+1}.$ The open neighborhoods for each odor source volume define open sets in the subspace topology. If we embed $R'$ by the canonical inclusion $\R^N\to \R^{N+1},$ then $R'$ is flat in $\R^{N+1}$ because the final coordinate of its elements is $0.$ Therefore, we can consider transformations of $R'$ that smoothly vary the final coordinate. For each transformation $f$, denote the resulting space as $S:=S(f)$; this constitutes the evolving perceptual space. Define the map $\Delta:R'\to S$ as the distension of $R'$ in $N+1$ (Diagram 1). This map arises from considering $M$ and $R'$ simultaneously, and is a diffeomorphism trivially.

 To better understand the map $\Delta,$ we here construct it as the composition of maps among the spaces already described, specifically showing how the (acquired) properties of $M$  govern the mapping of $R'$ to $S$. The map $B:R\to R'$ reflects glomerular-layer transformations as described above. For a fixed smooth section $\xi:R'\to M$ (which always exists by the triviality of $M$), we generate Diagram 2 (an elaboration of Diagram 1),

\begin{equation}\label{Diagram2} \begin{tikzcd}
R \arrow[r, "B"]  & R' \arrow[d,swap, "\Delta(f)"] \arrow[dr,"\xi"]&\\
           & S  &M \arrow[l,"\text{id}_{R'}\times f"] 
\end{tikzcd} \end{equation}
where $\Delta(f)$ is defined to be a map that makes the diagram commute. 
\begin{definition}
    Let $(x,U_x)$ be an odor source volume in $R'.$ We denote the image of this volume in $S$ as $(x,\widetilde{U_x}).$ This image denotes an $\textit{odor representation}$, also referred to simply as an $\textit{odor}$.
\end{definition}

\subsection*{The theory of sheaves enables a geometric framework based on local plasticity}
Plasticity in neural systems in general, and in the olfactory bulb in particular, is locally governed.  Changes in cellular and synaptic properties rely on the interactions of directly connected neurons and the locally regulated release of neurochemicals. These local effects, coordinated by sophisticated circuit interactions, collectively generate global performance at the network level. For example, in the olfactory bulb, the activity of mitral cells is shaped by local inhibition delivered by granule cells (Fig. 1). The action of a particular granule cell upon a particular mitral cell is uniquely determined via local synaptic plasticity, and it is this very uniqueness of each individual granule-to-mitral action that determines (collectively) the global transformation underlying perceptual learning.  However, while these local distortions transform the perceptual space (via distensions in $S$), they are neither globally governed nor constrained by the original geometry of the global system. To formally glue these many local plasticity operations, together with any relevant global processes, into a single analytical framework, we employ the theory of sheaves  \cite{Wedhorn2016} (Fig. 3). 

Briefly, we can consider olfactory bulb plasticity to modify local functions from $M-space$ to $S-space$.  Because this plasticity affects local neighborhoods consistently, the overlaps between these neighborhoods will agree, and then the collection of functions from $M$ to $S$ will form a sheaf. That is, the theory of sheaves enables us to glue all of these local functions together into a unique global function $M \to S$.  A more formal description follows.

\begin{definition} 
    A \textbf{presheaf} on a topological space $X$ with values in \textbf{Set} is a functor $F:\operatorname{Open}(X)^{op}\to \textbf{Set}.$  A \textbf{sheaf} is a presheaf that satisfies the additional conditions:
        \begin{enumerate}
            \item[(Sep)] If $\{U_i\}$ is an open cover of an open set $U,$ and there exist $s,t\in F(U)$ such that $s|_{U_i}=t|_{U_i}$ for all $i,$ then $s=t.$   
            \item[(Sh)] If $\{U_i\}$ is an open cover of an open set $U,$ and there exists $f_i\in F(U_i)$ for all $i$ such that $f_i|_{U_i\cap U_j}=f_j|_{U_i\cap U_j}$ for all pairs $(i,j)$ then there exists $f\in F(U)$ such that $f|_{U_i}=f_i.$ 
        \end{enumerate}
\end{definition}   

Presheaves satisfying the first property are referred to as \textit{separated} (the separation axiom implies that the second property yields unique lifting). To illustrate, consider the 
square blue section $\pi^{-1}(U_\alpha)$ of the M\"obius band in Figure 3. By covering the arc of the circle corresponding to this section ($U_\alpha$) with small segments $U_i$ (which correspond to vertical bars $\pi^{-1}(U_i)$ on the M\"obius band in the same way as the segment $p$ corresponds to the bar $\pi^{-1}(p)$), we then can consider continuous functions from the arc of small segments back into the square.  We denote this collection of functions $\Gamma(U,M)$. Corresponding to the open cover, we can consider and modify different collections $\Gamma(U_i,M)$ that are nonoverlapping subsets of $\Gamma(U,M)$.  We then pick functions $f_i$ on each $U_i$ such that they agree on the overlap of the sets. We then know we can construct a function $f:U_\alpha\to U_\alpha\times \R$ which restricts correctly. In this example case, the functions are simple, so the use of sheaves contributes little utility; however, when working with values that are more arbitrarily determined, sheaves enable them to be treated as a geometric whole. Indeed, sheaves can be considered generalizations of the M\"obius band (a vector bundle) in that the stalks/fibres $\pi^{-1}(p)$ are no longer required to be uniform.

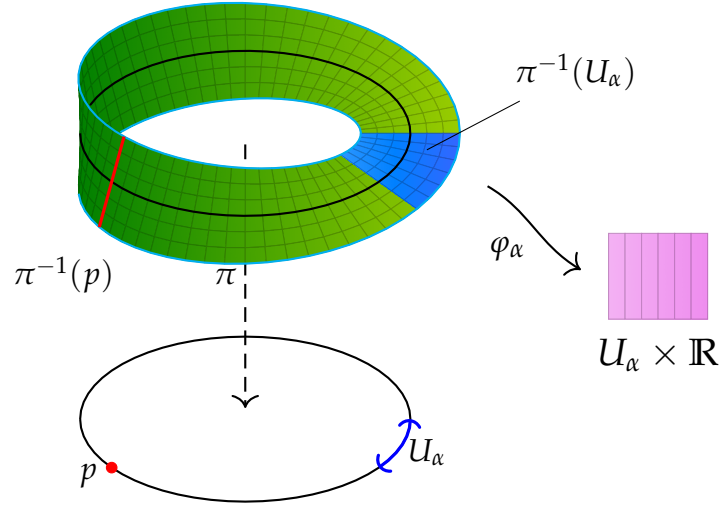
\begin{figure}     
\centering 
\begin{tikzpicture}[commutative diagrams/every diagram]
\centering
	\begin{axis}[axis equal, enlargelimits=false,
			hide axis,clip=false,zmax=1,zmin=-0.2, 
    view = {0}{30}
  ]
  \node (P2) at (axis cs: 0,0,-2) {};
  \node (P3) at (axis cs: 0,0,0) {};
  \path[commutative diagrams/.cd, every arrow, every label,thick,dashed] (P3) edge node[swap] {\Large $\pi$} (P2);
  
  \addplot3 [
    surf,
    colormap/greenyellow,
    shader     = faceted interp,
    point meta = x,
    samples    = 55,
    samples y  = 7,
    z buffer   = sort,
    domain     = 0:324,
    y domain   =-0.3:0.3
  ] (
    {(1+y*cos(x/2))*cos(x)},
    {(1+y*cos(x/2))*sin(x)},
    {y*sin(x/2)}
  );
  
  \addplot3 [
    surf,
    colormap/cool,
    shader     = faceted interp,
    point meta = x,
    samples    = 7,
    samples y  = 7,
    z buffer   = sort,
    domain     = 324:360,
    y domain   =-0.3:0.3
  ] (
    {(1+y*cos(x/2))*cos(x)},
    {(1+y*cos(x/2))*sin(x)},
    {y*sin(x/2)}
  );

  \addplot3 [
    surf,
    colormap/violet,
    shader     = faceted interp,
    point meta = x,
    samples    = 7,
    samples y  = 2,
    z buffer   = sort,
    domain     = -0.3:0.3,
    y domain   =-0.3:0.3
  ] (
    {2.5+x},
    {0},
    {-1+y}
  );

  \node at (axis cs:2.5,0,-1.55) {\large $U_\alpha\times \R$};
  
  \addplot3 [
    samples=58,
    domain=-180:160, 
    samples y=0,
    thick
  ] (
    {cos(x)},
    {sin(x)},
    {0}
  );

  \addplot3 [samples=113,domain=-168:506,samples y=0,thick,cyan] ({(1+0.3*cos(x/2))*cos(x)},{(1+0.3*cos(x/2))*sin(x)},{0.3*sin(x/2)});
 
  \addplot3 [
    samples=61,
    domain=-180:180, 
    samples y=1,
    thick
  ] (
    {cos(x)},
    {sin(x)},
    {-2}
  );

  \addplot3[domain=-0.3:0.3,samples=9,red,very thick,samples y=0]({(1+x*cos(216/2))*cos(216)},{(1+x*cos(216/2))*sin(216)},{x*sin(216/2)});
  \addplot3[(-),domain=324:360,samples=7,blue,very thick,samples y=0]({cos(x)},{sin(x)},{-2});
  \filldraw[red] (axis cs: -0.809,-0.5878, -2) circle[radius=2pt];
  \node at (axis cs: -0.95,-0.68,-2) {$p$};
  \node at (axis cs: 1.1,-0.4,-2) {$U_\alpha$};
  \node (P5) at (axis cs: -1.1, -0.5,-0.7) {$\pi^{-1}(p)$};
  \node (P4) at (axis cs: 2, 0.4,0.2) {$\pi^{-1}(U_\alpha)$};
  \draw (P4) -- (axis cs: 1.1,-0.2,0);
  \node (P0) at (axis cs: 1.4,-0.39,-0.1) {};
  \node (P1) at (axis cs: 2.1,0,-1.) {};
  \path[>=stealth,commutative diagrams/.cd, every arrow, every label,thick] (P0) edge[out=-30,in=150] node[swap] {\Large $\varphi_\alpha$} (P1);
  \pgfplotsset{compat=1.6}   
\end{axis}
\end{tikzpicture}

\vspace{0.5cm}
\caption{{\bf Illustration of the theory of sheaves.} The figure depicts a non-trivial vector bundle (a M\"obius band) over the circle. The projection map $\pi$ sends points on the M\"obius band to the corresponding point on the central circle. We therefore can pick coordinates on the M\"obius band such that, for each point $p$ in the circle, there is a line $\pi^{-1}(p)$ (the preimage, \textit{red}) in the band. The map $\varphi_\alpha$ is a local trivialization of the bundle on the open set $U_\alpha$, showing how the M\"obius strip is locally "flat" (whereas globally it is not). To obtain a sheaf from this picture, we consider the set of all possible embeddings of the circle (or segments thereof) into the M\"obius band. This will assign to a segment (such as $U_\alpha$) the collection of all such embeddings of this segment into $U_\alpha\times \R$. In this sense we think of the M\"obius band not as its own space, but instead as a collection of lines glued together in a particular way. This set of embeddings is a prototypical example of a sheaf. Modified from Figure 18 in \cite{Malek2018article} using code adapted from http://theoreticalphysics.info.} 
\vspace{0.5cm}
\hrule
\label{SheafDiag}
\end{figure}

Formally, we consider the local actions of granule cells onto mitral cells, and their concomitant modification of mitral cell output, as follows, considering that these actions may rely both on afferent sensory information and on additional inputs delivered onto granule cells from other sources such as piriform cortex \cite{Gao2009,Strowbridge2009}.  Recall from the previous section that for any smooth vector bundle $\pi:E\to P,$ we get two sheaves $C^\infty(-)$ and $ \Gamma(-,E)$ on $P$ such that $\Gamma(U,E)$ comes equipped with an action of $C^\infty(U)$ for all open $U\subseteq P.$  We here formally define an analogous pairing of sheaves to describe the modification by granule cells of afferent information contained in the mitral cell ensemble.

The first step in this formal definition is to define a functor \[ \mu:\mathcal{T} \to \textbf{R}^m\]
where $\mathcal{T}$ is the category defined by the topology on $R',$ and $\textbf{R}^m$ is the category whose objects are linear subspaces of $\R^m$ and the morphisms \[ \text{Mor}_{\textbf{R}^m}(U,V)=\begin{cases} \varnothing & U\not\subseteq V\\
\{*\} & U\subseteq V
\end{cases}\]  

\begin{lemma} 
    For any subspace $U\subseteq \R^m$, define a \textbf{sieve}  $\mathcal{S}$ on $U$ by a family of subspaces $V\subseteq U$ such that if $V'\subseteq V\in \mathcal{S}$ then $V'\in \mathcal{S}.$  Set $J(U)$ to be the collection of all such sieves on $U.$ Then, by this notion of sieve, $(\textbf{R}^m,J)$ defines a \textbf{site} (that is, $J$ defines a Grothendieck topology).   
\end{lemma} 
\begin{proof} 
    This lemma follows immediately from the proof that the category $\mathcal{T}$ with the standard notion of covering is a site. For such a proof, see \cite[Chapter III]{MacLaneMoerdijk1992}. 
\end{proof} 

Importantly, odorant presentations do not excite all receptor types, and therefore will activate some, but not all, mitral cells. This corresponds to the situation where $\xi(s)=(s,v)$ and $v$ has some coordinates equal to $0.$ The non-zero coordinates form a basis for some subspace of $\R^m.$ Let $n(\xi,s)$ be the number of non-zero coordinates in $v.$ Let $O$ be any open subset of $R'.$ Then \[\mu(O)=\R^\ell\]
where $\ell=\max\{n(\xi,p): p\in O\}.$ This construction shows that $\mu$ is functorial. 

\begin{lemma} 
    Let $\mathcal{F}$ be a sheaf on $\textbf{R}^m.$ Then, the functor $\mu^*\mathcal{F}(-):=\mathcal{F}(\mu(-))\in \textbf{Sh}(R').$
\end{lemma} 
\begin{proof} 
    The fact that this is a presheaf is immediate; thus it suffices to show that the gluing condition is satisfied. Let $U\subseteq R'$ be an open set and $\{U_i\}_{i\in I}$ an open cover of $U.$ Suppose further that we are given $f_i\in \mu^*\mathcal{F}(U_i)$ for each $i$ so that $\Res^{U_i}_{U_i\cap U_j} f_i=\Res^{U_j}_{U_i\cap U_j} f_j.$ We want to show that there exists some $f\in \mu^*\mathcal{F}(U)$ such that $\Res^U_{U_i}f=f_i.$  As $\{U_i\}$ is an open cover of $U,$ $\{\mu(U_i)\}$ will form a covering for $\mu(U)$ in the sense of a Grothendieck topology. Combining this with the fact that $\mathcal{F}$ is a sheaf on $\textbf{R}^m$ implies that there exists a unique $f\in \mu^*\mathcal{F}(U)$ satisfying the condition above. Hence, $\mu^*\mathcal{F}\in \textbf{Sh}(R').$   
\end{proof} 

\begin{lemma} 
    By abuse of notation, $C^\infty$ is a sheaf on $\textbf{R}^m.$ 
\end{lemma} 
\begin{proof} 
    The canonical functor $j_X:\operatorname{Open}(X)\to \textbf{Top}$ (which sends each open set $U\subset X$ to itself treated as a topological space and the inclusions $i_U:U\to X$ sent to the associated embeddings) induces a functor between the presheaf categories \[ j^*_X:\textbf{Set}^{\textbf{Top}^{op}}\to \textbf{Set}^{\operatorname{Open}(X)^{op}}\]given concretely by \[ j_X^*\mathcal{F}(U):=\mathcal{F}(U).\] Therefore, it follows that a presheaf on a category of topological spaces is a sheaf if and only if it is a sheaf on each topological space. As $C^\infty$ is a sheaf on each $\R^\ell,$ it follows that $C^\infty$ (by abuse of notation) is a sheaf on $\textbf{R}^m.$ 
\end{proof} 

\begin{corollary} 
    $\mu^*C^\infty\in \textbf{Sh}(R').$
\end{corollary}

All of this together provides a mathematical formalism for learning local data in $R'.$ We begin with a (small) open neighborhood of an odor $(x,U_x).$ Applying $\mu^*C^\infty$ we obtain the mitral cells with non-zero activation on points in this open set (and hence for all points in $U_x$) and the collection of all smooth functions on their output (the particular $\R^\ell$). The choice of a particular smooth function is then quantifying a local change in $S$.  Now, we define $G(-)$ as a flabby (flasque) sheaf of rings on $R'$ which act on $\mu^*C^\infty.$  This action performed by the sheaf $G$ is precisely the delivery of local inhibition onto mitral cells, and in particular to those mitral cells that are activated by a given odorant stimulus. This corresponds to the process by which activated receptors (fibers over the open set of the odor stimulus that are nonzero) propagate activity through the network based on the existing synaptic graph, and thereby induce plasticity according to local rules, without any need to globally update a learning map.  

With the definitions of $R,R',M, S,$ and $G$, we now can generate a fully explicit depiction of the model introduced in Diagram 1.  Using the theory of sheaves, we glue the local inhibitory actions encoded in sheaf $G$ into a global action, $\mu^*C^\infty$, that embodies the local-to-global transformations of these granule cell actions and enables their interaction with the global mapping of $M\to S$ that preserves $R'.$ Together, these underlie the concerted global transformation of the perceptual space -- i.e., the construction of $S$.

\begin{equation}\label{Diagram3} \begin{tikzcd}
R \arrow[r, "B"] & {(R',G,\mu^*C^\infty)} \arrow[rd, "\xi"] \arrow[d, "\Delta(g) "', dotted] &\\
& S  & M=R'\times \mathbb{R}^m \arrow[l, "\text{Id}_{R'} \times g"]
\end{tikzcd}\end{equation}

Formally, this diagram encompasses a commuting diagram of smooth manifolds. As we are considering sheaves on $R',$ all of these maps are indeed morphisms of ringed spaces. Newly associated with the space $R'$ are $G$, a sheaf constituting the granule cell modifications to perceptual output (which embeds the organism's prior learning), and the $G$-module $\mu^*C^\infty$, a sheaf of modules over each set of sections. As with Diagrams 1 and 2, we can track the path of a single odorant stimulus (a point in $R$) through the diagram until is it realized as a perceptual quality (a point in $S$). Specifically, an individual odorant is sent from $R$ to $R'$ via the map $B$, a set of signal conditioning transformations performed by glomerular-layer circuitry. $\xi$ then represents a particular choice of section of the vector bundle $M$ which immerses $R'$ into $M.$ This explicitly realizes the associations between glomeruli and mitral cells. The final morphism $\operatorname{Id}_{R'}\times g$ represents the building of the perceptual space. In total, $(\operatorname{Id}_{R'}\times g)\circ \xi$ is a diffeomorphism of $R'$ to $S$ in such a way that the modifications arising from perceptual learning are realized as increased heights in the final coordinates of points in the $N+1th$ dimension of $S$. The summary $\Delta(g)$, then, is simply the induced map by the composition. 

In sum, Diagram 3 illustrates how the theory of sheaves enables the formal representation of idiosyncratic, experience-dependent, locally-governed transformations of olfactory perceptual space within a global geometric framework. Odor learning generates locally-determined curvature within $S$-space, mapped as distensions into the $N+1$ dimension -- a transformation that has has several critical consequences.  First, these plastic changes underlie a process of category learning based upon the profiles of these quasi-discrete distensions, which correspond to potentially meaningful odor sources, reflecting the environmental reality that odor sources of interest are generally discrete but inclusive of natural variance.  Second, the preservation of physical similarity information within and among categories preserves a broad capacity to revise or elaborate learned categories, or even to remap similarity relationships based on new experience.  Third, as a necessary consequence of local plasticity, geometric frameworks for olfaction that are based on fixed curvature, whether it be Euclidean, hyperbolic, or otherwise, are unambiguously ruled out.  

\subsection*{Dynamics of odor category learning}

In the present framework, the process of odor learning generates a distension into the $N+1$ dimension of $S$ that reflects the odor source being learned, inclusive of variance.  This distension is an incipient category, and can be mapped back to an odor source volume in $R$ (Fig. 2A).  Olfactory perceptual learning studies suggest that such distensions are initially broad across N dimensions, but with accumulated experience come to reflect the actual N-dimensional quality variance profile exhibited by the natural odor source \cite{Cleland2009,Cleland2011,Moser_Bizo_Brown_2019}. The learned variances associated with each dimension of $R$ are presumed to be independent (receptor-specific), and the generalization gradients observed along any arbitrary trajectory through $R$ reflect the degree of distension into $N+1$ exhibited at each measured point along that trajectory.  

Critically, all of these transformations have timescales.  Learning takes time, and experiences that are rare, weakly attended, and/or of little consequence are not retained for long.  It is well established that fear memory persistence in the hippocampus depends strongly on the intensity of reinforcement, mediated by specific signaling pathways \cite{bekinschtein2007,bekinschtein2008}; similar mechanisms for memory regulation have been identified throughout the brain \cite{Dudai2015,Bellfy2020}, including within olfactory bulb \cite{Tong2014}.  For example, in olfactory habituation and spontaneous discrimination studies, in which odors are presented without an associated contingency, rodents' memory for these odors persists only on the order of minutes \cite{Hackett2015,Buseck2020,Freedman2013}. In contrast, when specific odors cue the availability of reward (reinforcement learning), odor memory after just 20 massed presentations is maintained for over two days (this persistence requires intact noradrenergic signaling and protein synthesis-dependent long-term memory consolidation mechanisms within the olfactory bulb) \cite{Tong2018,Linster2020}. Odors that are experienced as consistently meaningful over extended periods of time, of course, yield famously persistent memories \cite{Proust1913}.  Accordingly, the $S$-space landscape should be considered dynamic, with distensions into $N+1$ extending and often retracting in time according to the diverse underlying timescales of network plasticity.  Over time, however, the $S$-space of a given individual will develop a persistent topography in the $N+1$ dimension reflecting their accumulated knowledge and experience.  Representations of environmental odor sources, such as "orange", will be common to different individuals only by virtue of their shared experience.  

\subsection*{Discrimination learning}

Introducing curvature into $S$-space is fundamentally a process of unsupervised statistical learning, inclusive of reinforcement and/or other relevant influences.  One- or few-trial learning experiences generate broad generalization gradients \cite{Cleland2009,Cleland2011}, the breadths of which reflect the higher level of uncertainty afforded by undersampling. Consequently, responses to highly similar odors initially will tend to fall into the same consequential region; that is, the implicit interpretation would be that these odors are examples of within-category variance.  However, this changes when specific discrimination training paradigms are used.  Odor pairs that are not spontaneously distinguished can become rapidly distinguishable after they are associated with different reinforcers \cite{Linster2002}. While it is possible that discrimination between such similar odors could be achieved by asymptotic levels of statistical learning -- a gradual sharpening of their respective $S$-distensions into mature forms with reduced overlap -- the rapidity of discrimination learning and its dependence on reinforcement suggests an additional, directed process.  

In category learning, perceptual differences arise from increasing the intervening perceptual distance between categories (\textit{between-category separation}; \cite{PGJ-Harnad2019}).  Discrimination training is capable of rapidly and strongly rendering similar odors more perceptually different from one another than they were prior to learning -- that is, of selectively increasing the arc-length distance between their category representations in $S$. Consequently, to avoid an arbitrary floor effect constraint, discrimination learning must be able to not only retract between-category $S$-distensions to zero, but to extend them in the negative direction when warranted (Fig. 4B). A second difference between discrimination learning and ordinary statistical learning is that the former is specifically targeted \textit{between} two or more categories, rather than transforming the full extent of either individual category (Fig. 2C).  Accordingly, these retractions in $S$ will be localized between the specific categories that are being discriminated, and furthermore they may be emphasized specifically in those dimensions of $S$ in which the categories would otherwise overlap. Consequently, in the higher-dimensional $S$-spaces typical of biological olfactory systems, such localized retractions between two odors need not generate inappropriate side effects on the pairwise similarities between other odors.  Overall, discrimination learning serves as a prominent substrate of olfactory \textit{expertise}, in which trained individuals can easily and reliably recognize subtle distinctions that the untrained may not even perceive.  

Formally, consider two physically similar odorants $s^*=(x,\widetilde{U_x})$ and $t^*=(y,\widetilde{U_y})$ in $S.$  Because the early stages of odor learning are characterized by broadened generalization gradients \cite{Cleland2009}, reflecting sampling uncertainty, their odor representations (distensions in $S$) at this stage are likely to overlap:  $\widetilde{U_x}\cap \widetilde{U_y}\neq \varnothing.$  This is appropriate, given the prior likelihood that two highly similar odor stimuli, sampled in close succession, simply constitute two samples from the same odor source volume. To increase the discriminability of these similar odors, we construct a map that reduces only those values of $f$ which are sufficiently close (within some small $\varepsilon>0$) to a distance-minimizing path $\gamma$  connecting $x$ and $y$. Its existence follows from the existence of smooth bump functions on $M.$  Fix $f\in C^{\infty}(\R^m)$ so that $S=S(f).$ We consider functions $\alpha\in C^\infty(\R).$ Then, by defining the learning operation as $S\mapsto S(\alpha \comp f)$ we have a realization of this transformation by which two odor representations are progressively separated by learning. Denoting by $\widetilde{C^\infty(\R)}$ the constant sheaf corresponding to $C^\infty(\R)$ on $R'$, we here have defined a $\widetilde{C^\infty(\R)}$-module structure on $\mu^*C^\infty.$ Therefore, by considering only the interaction of $\alpha$ and $f$ over $\gamma,$ we have reduced the problem of discrimination learning to a one-dimensional problem. Accordingly, the map resulting from discrimination learning lengthens the perceptual metric $d^{per}$ between two similar odor source volumes, partitioning and expanding the previously shared space between the two representations so as to arbitrarily increase their perceptual dissimilarity without altering the physical distance $d_{phys}$ between their centers.
    
\begin{remark}
 	Based on the construction above, we can take $\widetilde{C^\infty(\R)}$ to be a rough approximation of $G$ as a sheaf. We cannot conclude that they are precisely equal, as this would require further analysis that is not presented here. 
\end{remark}

\begin{remark}
    Retractions in the $N+1$ dimension of $S$ owing to discrimination training between specific categories can result in violations of the triangle inequality \cite{TverskyGati1982} -- that is, discrimination training between odors A and B, but not between either of these and odor C, could lead to $d^{per}(ACB) < d^{per}(AB)$. This is psychophysically correct, but renders the perceptual metric $d^{per}$ not formally a metric. We nevertheless refer herein to $d^{per}$ as a metric for simplicity, and because this technicality does not affect how this measure of distance is used.  Interestingly, this recalls the argument by Tversky \cite{Tversky1977} that perceptual similarity, for this very reason, should be modeled using set theory rather than metric spaces.  We address this issue in the \textit{Semantic similarity and contrast models} section below.
\end{remark}

In summary, the process of discrimination training allocates resources to the specific categorical separation of odor sources that are highly physically similar and yet must be reliably distinguished.  This capacity to learn to perceptually distinguish very similar odor sources may underlie natural feats such as, for example, the odor-based social recognition of individual conspecifics \cite{Roberts2018,Thoss2015}. However, this acquisition of expertise is an effortful process, likely constrained by broader systemic limits on representational capacity, and hence cannot be casually extrapolated.  (Recent claims that humans can distinguish one trillion odors \cite{Bushdid2014} arise from this error, among other errors previously identified  \cite{Gerkin2015,Meister2015}). The development, over an individual's lifespan, of a richly featured landscape in $S$ is diagnostic of that individual's acquired olfactory expertise -- the richest examples of which may be found associated with hunter-gatherer subsistence strategies \cite{Majid2018} and with professions such as perfumer, chef, and sommelier.

\subsection*{Constructed hierarchical categorization}

Experience-dependent plasticity introduces curvature into $S$, distending and shaping incipient odor source categories and augmenting learned distinctions among them.  This process powerfully reshapes perceptual similarity space, modifying the perceptual distances $d^{per}$ among categorical odor representations without altering the physical distance metric $d_{phys}$.  We refer to the more persistent effects of learned experience as \textit{expertise}, and describe how this training enables an expert to easily distinguish subtly different odors that are perceptually indistinguishable to the untrained. However, experts also remain able to recognize the broader categories; being able to make fine distinctions does not imply that those distinctions are always appropriate or relevant to the task at hand.  Moreover, the finest distinctions are likely to require correspondingly higher-fidelity information, whether based on a stronger, cleaner signal, the accumulation of certainty over time, or both.  Notably, the olfactory system exhibits a speed-precision tradeoff \cite{Lahiri2016} (often conceived as a speed-\textit{accuracy} tradeoff owing to the experimental methods used to measure it), in which more difficult olfactory discriminations require correspondingly more time to resolve \cite{Abraham2004,Rinberg2006,Bhattacharjee2019,Frederick2017,Zariwala2013}.  All of these circumstances -- task-specific generalization, low-fidelity sampling, and time-constrained decisions based on partial information -- require the identification of unambiguous broader equivalence classes.  That is, if there is not enough certainty to identify a particular Walla Walla Cabernet Sauvignon, there may yet be enough to identify a Cabernet, or simply the scent of wine.  In this way, olfactory categorization can be conceived as hierarchical.  

Hierarchical categorical perception \cite{Bruni2008} is a natural emergent property of $S$-space curvature.  That is, based simply on the cumulative effects of statistical and discrimination learning, odor categories can admit more specific subcategories to an indefinite degree.  Lifelong experience thereby generates a mature landscape in the $N+1$ dimension of $S$-space, with greater olfactory expertise yielding a correspondingly richer, hierarchical landscape, studded with complex patterns of peaks, ridges, and valleys.  In a depiction of a mature $S$-space (Fig. 4), particular points in $R$ map directly onto the finest categorical resolution constructed by that individual.  The speed-precision tradeoff can be depicted as a gradual, progressive elevation of a sample point along the $N+1$ dimension of $S$, such that at any given time it may only be possible to identify a broader equivalence class within which the point is located, even if more time would permit a higher-resolution classification (Fig. 4A, \textit{arrows}).  Irreducible uncertainty in stimulus sampling can be modeled by a finite surface with dimensionality equal to $R'$; in the one-dimensional example depicted, this corresponds to a horizontal line segment depicting the resulting sample uncertainty, which may prevent the resolution of that sample beyond a bifurcation in the hierarchy, thereby limiting sample identification to a broader equivalence class (Fig. 4C, \textit{arrow}). For example, a sample of wine in the kitchen might be readily identifiable by a sommelier as pinot noir, but could only be further resolved by that individual into vintage and producer in a chemically 'quieter' environment.  In all cases, the number and hierarchical complexity of odor representations in $S$-space depend on the acquired expertise of the individual.

\begin{figure}[!ht]
  \vspace{0.5cm}
  \centering
  \includegraphics[width=6.5in]{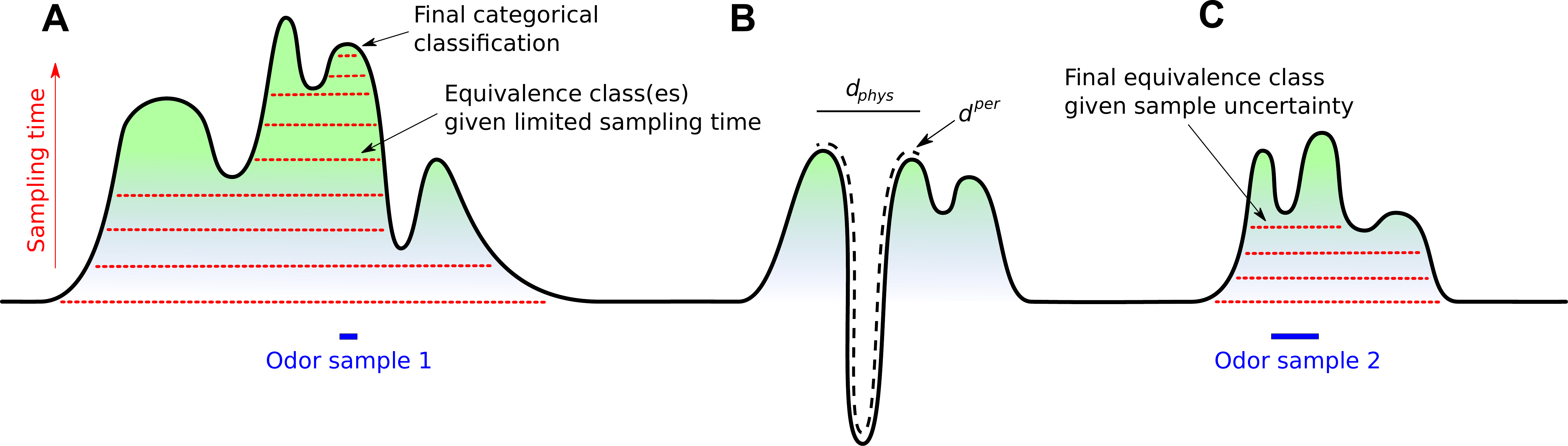}
  \vspace{0.25cm}
  \caption{{\bf Elaborated one-dimensional $S$-space with distensions into the $N+1$ dimension.}  Three broad categorical odor representations (A, B, C) are depicted, each admitting multiple narrower odor representations within the larger category. Odor samples (\textit{blue line segments}) exhibit a measure of irreducible sample uncertainty in \textit{d\textsubscript{phys}} that is denoted by the length of the segment (more generally, by a finite surface with dimensionality equal to that of $R'$). Odor sampling and recognition is depicted by migrating the odor sample progressively upward in the $N+1$ dimension on a behaviorally relevant timescale (\textit{sampling time} arrow).  During sampling, the internal representation of that odor sample at any given time will be the equivalence class corresponding to that point in time.  Accordingly, the initial representation (\textit{lowest horizontal dotted red line} in A and C) will be correspondingly broad, communicating only the identity of the broadest category to the animal.  With ongoing sampling, the non-irreducible uncertainty will be progressively reduced and the equivalence class will narrow into increasingly specific hierarchical subcategories as the sample progresses upwards in $N+1$ (\textit{horizontal dotted red lines}). This reflects the olfactory speed-precision tradeoff, in which odorants of greater physical similarity require correspondingly more time to reliably differentiate, whereas broader classification decisions can be made more rapidly if time constraints govern performance on a behavioral task \cite{Abraham2004,Rinberg2006,Bhattacharjee2019}. Importantly, discrimination learning increases \textit{d\textsuperscript{per}} specifically \textit{between} odor representations (\textit{dashed curve} in B, compare with \textit{d\textsubscript{phys}}), thereby enabling faster and more reliable discrimination between those representations (see also (Figure~\ref{Landscapes}C)). In high dimensional spaces, any two representations generally can be so separated without affecting the perceptual relationships of other neighboring odor representations.  That is, whereas in one dimension, as pictured above, the localized retraction between two specific odor representations within panel B also alters the perceptual distance between, for example, the odors depicted in panels A and C, in higher dimensional spaces this side effect would not follow.  Finally, odor samples with irreducible uncertainty that cannot be resolved into a single terminal hierarchical category can be ultimately classified at a lower hierarchical level (\textit{final equivalence class}, C).} 
  \vspace{0.5cm}
  \hrule
  \label{SSpace}
\end{figure}

\subsection*{Semantic similarity and contrast models}

In addition to distinct odors being perceived as similar based directly on their proximity in $S$, judgments of similarity and even odor identification can depend on context and other situational priors.  How can these phenomena arise from a geometric framework based on the receptor-based physical similarity of $R$-space?  Indeed, for this very reason, Tversky and colleagues argued that geometric models and hierarchies are not dependable bases for perceptual similarity.  Instead, similarity was defined using a contrast model derived from set theory, in which the similarity of two percepts depended on the number of features that they shared \cite{Tversky1977,TverskyGati1982}. (Exactly what counted as a feature for this quantitative purpose remained undefined). This set-theoretic framework naturally allowed that similarity might depend on context, in that some features might be weighted more strongly than others and that this weighting might change.  Such processes can easily lead to violations of the triangle inequality for objects that differ on two or more separable dimensions of similarity \cite{TverskyGati1982}, which traditionally has been interpreted as an argument against geometric models of similarity.

In the present geometric framework, however, local plasticity can freely generate violations of the triangle inequality in \textit{d\textsuperscript{per}}.  Discrimination learning, for example, can arise from selectively reducing the weighting of features (ultimately, receptor activities) that are common to two similar odors, based on the emerging configural recognition of one of these trained odors. That is, owing to local plasticity and aided by high dimensionality, the present geometrical framework can admit this essentially nonmetric transformation of similarity relationships. Indeed, the embedding of arbitrary priors (here arising from experience-dependent local plasticity) in high-dimensional neural systems has been recognized as having properties akin to set theory, in that these priors comprise features that can be shared among stimuli \cite{Singer2016} -- recalling existing arguments that the dichotomy of geometric and contrast models is ultimately immaterial \cite{Edelman2012}. The retention of geometrical properties is of course valuable for analysis, as contrast models otherwise admit no clear basis for representing the physical properties that underlie odor category variance and demarcation.

The weighting of features also can be influenced by external inputs to the olfactory bulb.  It is well established that ascending inputs from the piriform cortex and other regions of the brain can selectively activate granule cells \cite{Strowbridge2009,Gao2009}, potentially biasing bulbar odor representations with top-down priors \cite{Imam2020}. In the present framework, such inputs could transiently alter the geometric properties of the sheaf $G$ (Diagram 3), enabling a remapping of the similarity relationships among certain learned odor categories. For example, a top-down focus on features of sweetness might increase the similarity of orange to apple and reduce its similarity to lemon, dynamically modifying $d^{per}$ relationships in order to emphasize or disregard selectively targeted features of dissimilarity.  In contrast, a top-down focus on features of citrus would exert a contrary effect.  Representations of external cues, such as spatial context, also can bias the interpretation of olfactory stimuli \cite{Aqrabawi2016,Levinson2020}.  In sum, whereas we propose that experience-dependent transformations of $S$-space include persistent physiological changes that instantiate a critical substrate for long-term odor memory within the circuitry of the olfactory bulb and its immediate targets \cite{Tong2018,Tong2014}, the process of olfactory perception also admits dynamic, task-specific mechanisms that likely modulate these circuits substantively. 

\subsection*{Efficient coding}

The intrinsic dimensionality of olfactory encoding has been a topic of contention.  In part, this is a red herring, as dimensionality is not a property of the olfactory system per se, but rather characterizes particular models of specific aspects or stages of representation.  However, there is an important bifurcation in the set of existing models.  The first type of model is based on the intrinsic high dimensionality of $R$-space, and focuses on problems of primary stimulus encoding and neural circuit processing motifs (notably, these can differ substantially from analogous motifs employed in other sensory systems that operate on lower-dimensional feature spaces \cite{Cleland2006,Cleland2010,Cleland2014}). Such models emphasize the coding potential and corresponding low-level physiological mechanisms of relatively peripheral layers of the olfactory system. In contrast, the second type of model is generally based on reports of psychophysical similarity, the results of which then are condensed onto a reduced dimensionality using tools like principal components analysis \cite{Koulakov2011, Zarzo2006} or non-negative matrix factorization \cite{Castro2013}.  These graph-theoretic models eschew questions of coding potential in favor of depicting the current perceptual relationships among odors that already have been encoded.  The specific dimensionality estimates in these works are necessarily limited by the scope of the odor sets employed, and certainly are influenced by the nature of the questions used to obtain the data, but this does not invalidate the basic concept that one can legitimately construct psychophysical spaces of substantially lower dimensionality than the $R$- and $S$-spaces discussed herein.  What is the research value of these descriptive spaces, and how do they relate to the present framework?  

The broadest arguments for reduced dimensionality are related to the efficiency of memory search -- i.e., the ability to quickly recognize known stimuli along with perceptually similar variants that may have similar implications.  High-dimensional spaces are vast, and the efficiency of retrieval is paramount for organisms in the wild.\footnote{The problem posed by high dimensionality to memory search efficiency may be overemphasized, because the control-flow methods used by contemporary computer architectures are particularly vulnerable to the curse of dimensionality. The decentralized architecture and localized computational tactics employed by brain circuits need not suffer from this curse.}  Fortunately, $S$-space is highly reducible, because the olfactory modality is \textit{signal sparse} \cite{Berke2017}.  This sparseness property is not directly related to the proportion of neurons that are active at any given time; rather, it indicates that, of all the distinguishable states that a neural system could assume, only a very few are ever actually occupied. That is, within a bewilderingly high-dimensional $R$-space, the vast majority of that volume corresponds to no meaningful odor and may never even be experienced in an animal's lifetime.  

An efficient system will adapt to signal sparseness, preferentially allocating resources to encode the properties of stimuli that actually exist \cite{Beyeler2019,Olshausen2004}.  Current models of bulbar plasticity achieve this by utilizing configural plasticity and the selective allocation of adult-generated interneurons (adult neurogenesis) to enable lifelong learning capacities within active, occupied regions of $R$-space \cite{Imam2020,Borthakur2019Front}. These regions of stimulus-occupied space can then, in principle, be selectively projected onto a new basis that is representationally compact and amenable to efficient search.  Two classes of transformation that enable such dimensionality reduction while preserving similarity relationships are those that enable \textit{categorization} and those that support \textit{regression} \cite{Edelman2012}.  The transformations of the present framework afford the advantages of both of these methods, enabling experience-dependent category formation while preserving a physical basis for intercategory similarity upon which further basis transformations can be performed.  For present purposes, this dual capacity enables the construction of transformed spaces comprised of the category representations of known odors, remapped onto a new basis that excludes unoccupied regions of $R$-space and may either directly inherit the similarity relationships of $d^{per}$ or modify them owing to additional higher-level inputs.  We refer to these transformed spaces as $T$-spaces.

Notably, operational $T$-spaces in this framework must be dynamic.  That is, if categorical odor representations are constructed through plasticity, then newly learned representations (or modifications of existing representations that acquire new qualities or $d^{per}$ relationships) must be able to establish themselves within the similarity relationships of $T$-space.  Much of the time, this enrichment will require some reconstruction of an existing $T$-space (potentially including a modest expansion of $T$-dimensionality), and therefore must induce a degree of ongoing remapping of the underlying physical representation across some postbulbar area(s) such as piriform cortex and olfactory tubercle (\textit{aka} tubular striatum \cite{Wesson2020}).  Indeed, odor representations in piriform cortex do appear to exhibit progressive remapping over time \cite{SchoonoverFink2021}.  Moreover, such representational drift increasingly appears to be the rule rather than the exception across the brain \cite{Rule2019,Rule2020}, being evident particularly in visual cortex \cite{Deitch2021}, parietal cortex \cite{Driscoll2017}, and hippocampus \cite{Ziv2013,Mankin2015}. Additional factors potentially affecting $T$-space formation include the merging of contextual information into odor representations, likely even within olfactory bulb \cite{Aqrabawi2016,Levinson2020,Aqrabawi2020}.  For present purposes, however, $T$-spaces remain weakly defined. They are not necessary elements in the present construction, except as an illustration of the theoretical reducibility of $S$. Indeed, postbulbar "odor representations" in the brain may be better described as odor-informed state or goal representations, in which $T$-spaces are dynamically constructed for cause and necessarily integrative and multimodal.  

\subsection*{Conclusion}

We present a general geometric framework outlining the operation of the vertebrate early olfactory system and illustrating how category learning underlies the efficient encoding of environmental odor sources into quasi-discrete, meaningful \textit{odors} and underlies the development of olfactory expertise.  The construction of the resulting $S$-space corresponds to experience-dependent plasticity within the circuitry of the deep olfactory bulb, inclusive of its reciprocal interactions with its immediate follower cortices.  Because of the heuristic, statistical, and distributed nature of this plasticity, a single geometry with fixed curvature does not suffice to describe the system; instead, the theory of sheaves is used to glue these diverse, localized changes into a coherent geometric whole.  The framework unifies a broad range of experimental results and theory, from the physics of ligand-receptor interactions, to the plasticity of bulbar circuits and the acquired configural receptive fields of interneurons, to the formation of generalization gradients, the speed-precision tradeoff, and the capacities and predictions of competing cognitive models of similarity.  Categorical odor perception is depicted as an ongoing, constructive process of segmentation arising from deformations of $S$-space, admitting multiple hierarchical layers within categories and multiple timescales of local plasticity.  This framework is sufficiently concrete to serve as a roadmap for the development of operational brain-inspired artificial systems capable of rapid prototype learning.  

\subsection*{Acknowledgments}
We gratefully acknowledge Alec J. Mutti and Zarina R. Lagman for assistance with figures, and the members of the Computational Physiology Laboratory for challenging discussions and advice.  

\pagebreak
\bibliographystyle{\style}
\bibliography{references}

\end{document}